\documentclass[aps,prapplied,reprint,superscriptaddress,longbibliography,nofootinbib,floatfix,raggedbottom]{revtex4-2}

\usepackage[T1]{fontenc}
\usepackage{lmodern}
\usepackage{amsmath,amssymb,bm}
\usepackage{graphicx}
\usepackage{booktabs}
\usepackage{multirow}
\usepackage{array}
\usepackage{xcolor}
\usepackage{etoolbox}
\usepackage{hyperref}
\hypersetup{colorlinks=true,linkcolor=blue!55!black,citecolor=blue!55!black,urlcolor=blue!55!black}

\AtBeginEnvironment{thebibliography}{\footnotesize\linespread{0.95}\selectfont}

\newcommand{\UF}{U_{\mathrm F}}
\newcommand{\Uin}{U_{\mathrm{in}}}
\newcommand{\Rreset}{\mathcal{R}_{p}}
\newcommand{\Ddeph}{\mathcal{D}_{\lambda}}
\newcommand{\rhozero}{\rho_{0}}

\begin{document}

\title{Coherent Floquet quantum reservoirs for molecular property prediction}

\author{Luofei Wang}
\affiliation{School of Physics, Beijing Institute of Technology, Beijing 100081, China}
\author{Da Zhang}
\affiliation{School of Physics, Beijing Institute of Technology, Beijing 100081, China}
\author{Congren Wang}
\affiliation{School of Physics, Beijing Institute of Technology, Beijing 100081, China}
\author{Yiming Li}
\affiliation{School of Physics, Beijing Institute of Technology, Beijing 100081, China}
\author{Yuxiao Yang}
\affiliation{School of Physics, Beijing Institute of Technology, Beijing 100081, China}
\author{Xuan Zhang}
\affiliation{School of Computer Science and Technology, Shandong University, Qingdao 266230, China}
\author{Xuefeng Cui}
\affiliation{School of Computer Science and Technology, Shandong University, Qingdao 266230, China}
\author{Zhang-Qi Yin}
\email{zqyin@bit.edu.cn}
\affiliation{School of Physics, Beijing Institute of Technology, Beijing 100081, China}

\date{\today}

\begin{abstract}
Quantum reservoir computing (QRC) uses observables of a driven quantum system to make predictions from a sequence of inputs. The robust subharmonic response of discrete time crystals (DTCs) has motivated their use as quantum reservoirs. Here we develop a DTC-QRC architecture for molecular property prediction in which successive graph events or surface-hopping frames drive a fixed Floquet reservoir. An adjustable reset probability controls how long earlier inputs influence the state, and measurements of the final polarizations and pair correlations yield a feature vector of fixed dimension. Weak reset favors classification of inhibitor activity and blood--brain barrier penetration from long graph prefixes, whereas forecasting electronic gaps in ethene favors greater weight on recent nuclear motion. With nonlinear decoding, DTC-QRC outperforms an echo-state network on these tasks at matched input lengths and readout widths. For gap forecasting with this decoder and final-state readout, the dissipative reservoirs share a low-error drive window near the DTC transition reported for related spin chains. Dephasing lowers prediction performance in both molecular applications. In six-qubit classification without programmed reset on the Quafu superconducting quantum cloud platform, pair correlations retain more of their noise-free predictive power toward the DTC side, and the full readout performs best near the transition among the drives tested. The architecture provides a common framework for molecular screening and time-resolved property prediction using quantum reservoir computing.
\end{abstract}

\maketitle
\raggedbottom
\clubpenalty=10000
\widowpenalty=10000

\section{Introduction}

Chemical groups separated by several bonds can contribute to the same molecular activity. A related problem arises in molecular dynamics, where successive observations of nuclear motion precede a change in electronic energy. Both problems can be expressed as prediction from a sequence of local observations. For a reservoir of fixed size, the question is how to combine those observations so that the final measurement retains the information needed to predict the property.

Local chemical representations already support molecular screening. DeepDTA learns from chemical strings and protein sequences to predict drug--target binding affinity \cite{Ozturk2018DeepDTA}, and graph-based models have guided the discovery of antibacterial molecules, including halicin \cite{Stokes2020Antibiotic}. When the input arrives one chemical event at a time, groups that contribute to the same prediction can enter at different steps. A sequential predictor must carry information about the first group forward until it encounters the second.

Molecular trajectories give the sequence a physical clock. VAMPnets, for example, learn kinetic models from time-lagged configurations and recover slow conformational processes such as protein folding \cite{Mardt2018VAMPnets}. For a forecast based on selected coordinates, two trajectories can reach similar geometries after following different paths. Earlier frames help distinguish these paths, but how far back a predictor should look depends on the property and the forecast horizon.

Reservoir computing processes such sequences with fixed dynamics and a trained decoder \cite{Maass2002,Jaeger2004,Tanaka2019}. In quantum reservoir computing (QRC), observables of a driven quantum system supply the decoder inputs \cite{Fujii2017,Nakajima2019,Ghosh2019,Mujal2021AQT,FujiiNakajima2021Book}. Studies of reservoir dynamics and dissipation have established how they affect memory and nonlinear response \cite{MartinezPena2021PRL,Bravo2022PRXQ,MartinezPena2023PRE,Sannia2024Quantum}. Complementary work addresses measurement, coherence, feedback, and decoding \cite{Mujal2023npj,Kubota2023PRR,Domingo2023SR,Fry2023SR,Cindrak2024PRR,Kobayashi2024PRE,Kobayashi2024PRXQ,Palacios2024CP}. A nine-spin experiment has compared weather prediction with classical reservoirs containing thousands of nodes \cite{Hou2026PRL}. For molecular prediction, the question is how the dynamics bring features of several local observations into the small set of observables available to a decoder.
Earlier work used a discrete-time-crystal (DTC) reservoir to classify images, encoding each sample once before Floquet evolution and readout \cite{ZhangLiGuoYuJinYin2025}. That work motivates our use of a DTC reservoir for molecules, where each new event acts on the state left by its predecessors.

With near-$\pi$ pulses, interactions and disorder can stabilize a subharmonic response against perturbations \cite{Khemani2016,Else2016,VKhemani2016PRB,Yao2017,NandkishoreHuse2015,Lazarides2015PRL,vonKeyserlingk2016Stability,Abanin2019RMP}. Experiments have explored this response in trapped ions, diamond spins, and superconducting qubits \cite{Zhang2017Nature,Choi2017Nature,Randall2021Science,Mi2022Nature}. For a related Floquet spin chain, Mi et al. observed a finite-system crossover near the predicted transition at $g\simeq0.84$ \cite{Mi2022Nature,Ippoliti2021PRXQ}. This supplies a physical reference for choosing a drive that can mix successive inputs while preserving information about them under noise.

Here we construct a DTC-QRC architecture in which each local observation sets the qubit rotations preceding a fixed number of Floquet cycles. The reset probability sets how long the influence of earlier events persists. At the end of the sequence, we measure single-qubit polarizations and pair correlations. Only the classical decoder is trained, and its input dimension stays fixed as the sequence grows.

For inhibitor-activity and blood--brain barrier permeability classification, breadth-first search (BFS) converts each molecular graph into local events that record atoms, bonds, ring closures, and traversal depth. A longer prefix exposes more of the structure of the same molecule. For electronic-gap prediction, we instead form windows from ethene surface-hopping trajectories in the SHNITSEL excited-state collection \cite{Curth2025SHNITSEL}. Here the observations are ordered in physical time, and the target is an energy gap after the last observed frame.

At matched input lengths and readout widths, DTC-QRC outperforms an echo-state network (ESN) on long graph prefixes and on the ethene gap forecasts. In simulation, the preferred reset probability depends on the molecular target. Weak reset allows chemical information from different parts of a graph to accumulate, whereas the gap forecasts benefit from giving recent motion greater weight. Even at complete reset, the drive affects how accurately the last encoded frame predicts the gap. For dissipative gap forecasting with nonlinear decoding and final-state readout, a common low-error window lies near the DTC transition reference. In BACE classification on the Quafu superconducting quantum cloud platform, second-order correlations retain more predictive information on the DTC side, while the full $Z+ZZ$ readout attains its highest mean classification performance near the transition among the sampled drives. The hardware comparison thus favors a drive that mixes successive inputs while preserving the correlations that distinguish active from inactive compounds.

\section{Model and methods}

\begin{figure*}[t]
\centering
\includegraphics[width=0.95\textwidth]{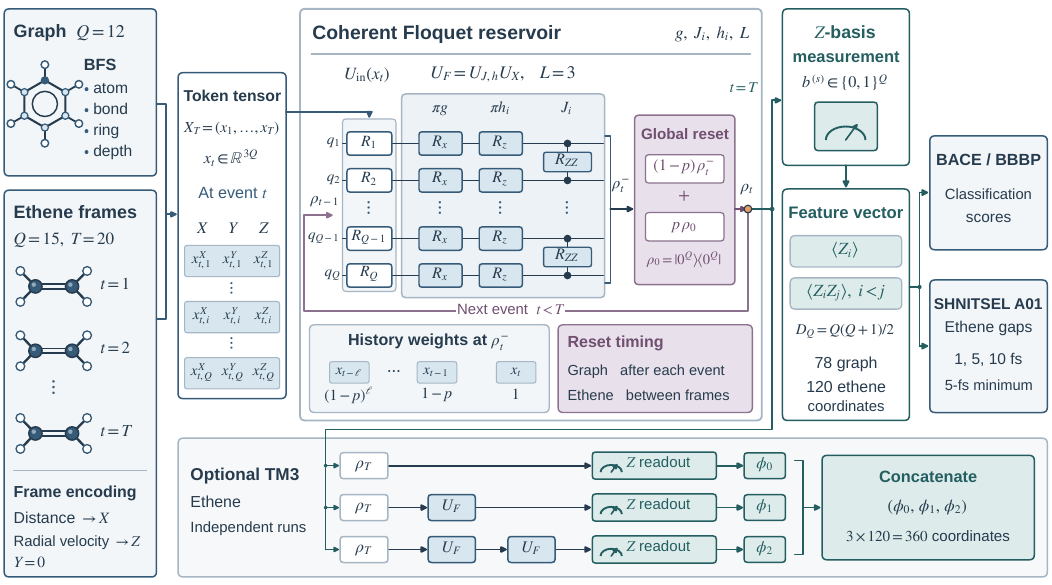}
\caption{Molecular property prediction with DTC-QRC. Local BFS events drive a 12-qubit reservoir for BACE inhibitor-activity and BBBP permeability classification. A 15-qubit reservoir encodes 20 ethene frames of interatomic distances and radial velocities to forecast $E_1-E_0$ at 1, 5, and 10 fs after the last frame and its minimum over the next 5 fs. Each event is encoded once before three Floquet cycles. Global reset acts with probability $p$ after each graph event or between trajectory frames. The two output arrows denote classical decoding of $Z+ZZ$ features. TM3 concatenates independent readouts after zero, one, and two extra cycles without further input or reset. The schematic was prepared with AI assistance.}
\label{fig:circuit}
\end{figure*}

The two tasks use different molecular inputs but share the same computational sequence. Each event sets a local rotation, Floquet evolution couples the qubits, and endpoint measurements supply the features for classical decoding (Fig.~\ref{fig:circuit}).

\subsection{Molecular event encoding}

For graph classification, we parse a canonical isomeric Simplified Molecular Input Line Entry System (SMILES) string into a molecular graph $G=(V,E)$ and traverse it by BFS \cite{Weininger1988}. We encode atom discovery, parent--child bonds, and ring closures together with traversal depth and padding. For a reservoir of $Q$ qubits, processing $T$ events gives the input sequence
\begin{equation}
 X_T(G)=\left(x_1,x_2,\ldots,x_T\right),\qquad x_t\in\mathbb{R}^{3Q}.
 \label{eq:graph_stream}
\end{equation}
We call this the \emph{local chemical-prior token} protocol. Each event fills a fixed ordering of local atom and bond attributes and traversal context, from which we retain the first $3Q$ entries. For example, a carbon-atom event with three bonded neighbors begins with $(1,0,0,6/80,3/6,0)$. The first triple identifies an atom event. The second contains the normalized atomic number and degree and a reserved zero. Consecutive triples supply the $X$, $Y$, and $Z$ Hamiltonian coefficients on each qubit in Eq.~(\ref{eq:hamiltonian_injection}).

The entries reserved for whole-molecule descriptors are fixed at zero, so the input contains only the local events and their traversal context. Reparsing the canonical SMILES fixes the atom indices before traversal and gives a deterministic event order. Increasing $T$ reveals more of the graph and increases circuit depth. BACE records $\beta$-secretase 1 inhibitor activity and BBBP records blood--brain barrier penetration. The processed datasets contain 1513 BACE records and 2039 BBBP records, corresponding to 1513 and 1975 unique canonical SMILES, respectively \cite{Wu2018MoleculeNet}. In all six paired Bemis--Murcko scaffold splits \cite{BemisMurcko1996}, repeated records of the same canonical structure stay in one partition. Appendix~\ref{app:protocol} gives the seed values and duplicate counts.

For time-resolved gap forecasting, we use SHNITSEL A01 ethene trajectories \cite{Curth2025SHNITSEL}. Each frame supplies 15 interatomic distances and their backward-difference radial velocities. We standardize each pair/channel using training trajectories and clip the standardized values to $[-5,5]$. Distance sets the $X$ coefficient and radial velocity sets $Z$, with $Y=0$. Both inputs carry a scale factor of 0.5. Each sample has $T=20$ events at 0.5-fs spacing. To compute the first velocity, we use one preceding position frame, so the raw observations span 10 fs. The four targets are the $E_1-E_0$ electronic energy gap 1, 5, and 10 fs after the endpoint and its minimum over the next 5 fs. The minimum captures brief passages through small gaps between forecast times, when nonadiabatic transitions can become important. Eight deterministic endpoints per trajectory give 2376 windows from 297 trajectories. A trajectory-disjoint split assigns 178, 59, and 60 trajectories to training, validation, and testing. The reservoir uses the same six seeds as the graph task.

In both tasks, the inputs set the local Hamiltonians, and we use the same Floquet rule and $Z+ZZ$ observables. Graph classification uses $Q=12$ and samples the final state after applying the reset channel at every event. Surface hopping uses $Q=15$, resets between frames, and compares final-state sampling with three-sample temporal multiplexing (TM3).

\subsection{Floquet dynamics and controlled reset}

We initialize all qubits in $\rhozero=|0^Q\rangle\langle0^Q|$. On qubit $i$, three real coordinates define the local Hamiltonian exponential,
\begin{equation}
 \Uin(x_t)=\bigotimes_{i=1}^{Q}
 \exp\!\left[-\frac{\mathrm{i}}{2}
 \left(x_{t,i}^{X}X_i+x_{t,i}^{Y}Y_i+x_{t,i}^{Z}Z_i\right)\right].
 \label{eq:hamiltonian_injection}
\end{equation}
For each qubit, the three coordinates specify a vector in spin space. Its direction sets the rotation axis and its norm sets the rotation angle, translating the molecular attributes into a local change of the quantum state. The tensor-product form injects each event through single-qubit operations. Interactions in the subsequent Floquet cycles couple these locally encoded components, so the measured correlations can depend on combinations of molecular attributes.

Graph events supply the three rotation coefficients directly. For a surface-hopping frame, the scaled distance and radial velocity place the rotation axis in the $XZ$ plane. In both cases the input fixes a single rotation on each qubit, while the couplings that act between inputs stay fixed throughout decoder training.

After each event, we apply $L$ Floquet cycles,
\begin{align}
 \rho_t^{-} &= \mathcal{F}_{x_t}(\rho_{t-1}),\nonumber\\
 \mathcal{F}_{x}(\rho) &=
 \left(\UF^{L}\Uin(x)\right)\rho
 \left(\UF^{L}\Uin(x)\right)^{\dagger},\nonumber\\
 \UF &= U_{J,h}(W)U_X(g),\nonumber\\
 U_X(g)&=\prod_i\exp[-\mathrm{i}g\pi X_i/2],\nonumber\\
 U_{J,h}(W)&=\exp\!\left[-\frac{\mathrm{i}}{2}
 \left(\sum_{i=1}^{Q-1}J_iZ_iZ_{i+1}
 +\pi\sum_{i=1}^{Q}h_iZ_i\right)\right].
 \label{eq:floquet}
\end{align}
The input rotation acts on the state left by the preceding event. Each Floquet cycle then applies the transverse pulse followed by the longitudinal fields and nearest-neighbor conditional phases. The rotations and interactions generally do not commute, so changing the order of two inputs can change the final state. Repeated cycles carry this dependence on the input sequence into the one- and two-body observables measured at the endpoint.

The reservoir seed fixes $J_i\sim\mathcal{U}[-0.9,0.9]$ and $h_i\sim\mathcal{U}[-W,W]$. Prior DTC and DTC-QRC studies motivate the fixed drive $g=0.84$ used in the application comparisons \cite{Mi2022Nature,ZhangLiGuoYuJinYin2025}. We set $W=0.50$ and $L=3$ throughout these simulations and examine the drive dependence in Sec.~\ref{sec:gscan}.

After graph event $t$, we apply the reset channel
\begin{equation}
 \rho_t=\Rreset(\rho_t^{-})=(1-p)\rho_t^{-}+p\rhozero.
 \label{eq:reset}
\end{equation}
For graph streams, reset follows every event, including the last. For surface hopping, it acts only between frames, leaving the final input step and any subsequent sampling cycles unitary. We evaluate the channel as an exact mixture of evolutions starting at different positions in the sequence. The closed reservoir uses $p=0$, and the graph application fixes $p=0.10$. For gap forecasting we select one $p$ for each target, sampling scheme, and decoder family by minimizing the validation mean absolute error (MAE) averaged over seeds.

To see how reset weights the history, we expand the graph recurrence. We define
$\mathcal{F}_{b:a}=\mathcal{F}_{x_b}\circ\cdots\circ\mathcal{F}_{x_a}$ and
$\mathcal{F}_{T:T+1}=\mathcal{I}$. Unrolling Eq.~(\ref{eq:reset}) gives
\begin{equation}
 \rho_T=(1-p)^T\mathcal{F}_{T:1}(\rhozero)
 +p\sum_{k=1}^{T}(1-p)^{T-k}
 \mathcal{F}_{T:k+1}(\rhozero).
 \label{eq:suffix_mixture}
\end{equation}
The first term is the history with no reset. In the $k$th summand the last reset follows event $k$, so only events after $k$ act on the newly prepared state. A history survives $\ell$ subsequent reset locations with probability $(1-p)^{\ell}$. For $0<p<1$ the corresponding decay length is $[-\ln(1-p)]^{-1}$ events. Thus $p$ sets the weights of histories of different lengths. What the decoder can recover from each history depends on the Floquet evolution and the measured operators. Reset between trajectory frames gives the same geometric weighting and leaves the last frame unreset.

\subsection{Readout and temporal multiplexing}

The final-state representation contains every one-body and pairwise computational-basis observable,
\begin{equation}
 \begin{aligned}
 \Phi_Q(\rho)&=\left[\{\langle Z_i\rangle\}_{i=1}^{Q},
 \{\langle Z_iZ_j\rangle\}_{i<j}\right],\\
 D_Q&=\frac{Q(Q+1)}{2}.
 \end{aligned}
 \label{eq:observables}
\end{equation}
A single set of computational-basis measurements estimates all $D_Q$ entries. Let $\mathcal{C}_{X_T}$ denote the complete input-dependent channel for one event stream and let $O_a\in\{Z_i,Z_iZ_j\}$. Each endpoint coordinate has equivalent Schr\"odinger and Heisenberg forms,
\begin{equation}
 \phi_a(X_T)=\operatorname{Tr}\!\left[O_a\mathcal{C}_{X_T}(\rhozero)\right]
 =\operatorname{Tr}\!\left[\mathcal{C}_{X_T}^{\dagger}(O_a)\rhozero\right].
 \label{eq:heisenberg_observable}
\end{equation}
Although $O_a$ acts on at most two qubits, evolving it backward through the channel in Eq.~(\ref{eq:heisenberg_observable}) makes its dependence on the ordered inputs explicit. Measuring in a single basis at the endpoint can therefore reveal relations among events that entered at different steps. We call such a relation \emph{readable} when a specified decoder can recover it from the measured coordinates.

For final-state sampling, the prediction follows
\begin{equation}
 \text{event history}\;\longrightarrow\;\rho_T
 \;\longrightarrow\;\Phi_Q(\rho_T)
 \;\longrightarrow\;\widehat{y}.
 \label{eq:readable_chain}
\end{equation}
Here $\widehat{y}$ denotes the prediction score or gap estimate.

Graph classification uses final-state sampling. Surface-hopping regression also uses three-sample temporal multiplexing (TM3), which concatenates $Z+ZZ$ at the endpoint and after one and two additional Floquet cycles. During these continuations, we apply neither new inputs nor reset. For $Q=15$, final-state sampling exposes 120 coordinates and TM3 exposes 360. Hardware acquisition of TM3 would require separate executions at the three sampling depths.

For a given input, reservoir, reset probability, and sampling protocol, the same measured coordinates can be used to fit decoders for different targets. Changing $p$ changes the representation. In simulation we recompute the features by reweighting the stored contributions from histories of different lengths. As more events enter the reservoir, the circuit becomes deeper while the final-state readout still has $D_Q$ coordinates.

\subsection{Classical decoding and baselines}

An affine decoder tests whether a weighted sum of the measured observables plus a bias suffices to predict the target. Radial-basis-function support-vector regression (RBF-SVR) can also use nonlinear relations among those observables. Fitting both decoders to the same coordinates shows how much predictive performance changes when the readout can use these nonlinear combinations.

We fit RBF-SVR to standardized coordinates and use its continuous output as the molecular activity or permeability score, evaluated by the area under the receiver-operating-characteristic curve (ROC-AUC, abbreviated AUC). For ethene, the output estimates the gap and we measure its mean absolute error. The affine controls use logistic regression for classification and ridge regression for gap forecasting. At each seed and reservoir setting, the validation partition selects the decoder. For ethene we then refit the selected model on the combined training and validation trajectories before testing. Appendix~\ref{app:protocol} gives the search grids and selection rules.

The ESN takes the same $T\times3Q$ input sequence as QRC. After applying $\tanh$ and a random input projection, it uses each event for three recurrent updates, whereas QRC injects that event once before three Floquet cycles. The two models share the data splits, seeds, and decoder grids. Both expose 78 coordinates for graph classification and, on the same 20-frame ethene windows, 120 coordinates at the endpoint or 360 with TM3. Appendix~\ref{app:esn_width} examines how the graph comparison changes with ESN width.

Full-graph message passing and chemical fingerprints predict properties from broader structural information \cite{Gilmer2017MPNN,Xu2019GIN,Rogers2010ECFP}. The matched ESN comparison asks how two reservoirs process the same sequence of local observations at the same readout width.

\subsection{Statistical analysis and dephasing}

Graph curves show six-seed means and sample standard deviations. At each reported $T$, we obtain a pointwise 95\% percentile interval by resampling the six paired QRC--ESN differences. For the crossed percentile intervals in gap forecasting, we resample the six seeds and 60 test trajectories independently, keeping the methods paired within each draw. At each reservoir setting we first select decoder hyperparameters by validation MAE within each seed. The application comparison uses the reset probability with the lowest six-seed mean validation MAE, with ties resolved in favor of smaller $p$. The reset-response curves show the uniform grid from $p=0$ to 1. Appendix~\ref{app:protocol} gives the resampling settings and additional low-$p$ candidates.

The graph reset scan fixes $Q=12$ and reports complete response curves for $T\in\{6,12,20,30\}$. For each reset value, we compare the result with that of the closed reservoir at the same $T$.

To introduce event-wise dephasing, we apply the channel
\begin{equation}
 \Ddeph=\bigotimes_{i=1}^{Q}
 \left[\left(1-\frac{\lambda}{2}\right)\mathcal{I}
 +\frac{\lambda}{2}\mathcal{Z}_i\right],
 \qquad \mathcal{Z}_i(\rho)=Z_i\rho Z_i,
 \label{eq:dephasing}
\end{equation}
after every event. We average $k=8$ or 32 Pauli trajectories per input sample and suffix at $\lambda=0.5$ and 1, with exact reset weights and $p$ fixed at 0 or 0.10. The $\lambda=0$ reference uses exact coherent evolution. BACE uses $Q=12$, $T=30$, while hopping uses $Q=15$, $T=20$. Both use final-state $Z+ZZ$ sampling and RBF-SVR. Each dephasing condition follows the original split and decoder-selection rules. For hopping, we divide each target MAE by its coherent value within seed and physical test trajectory at the same $p$, then average over four targets and 60 trajectories.

\section{Molecular graph classification}

For graph classification, each molecule contributes a prefix of $T$ BFS events and one label, inhibitor activity for BACE or blood--brain barrier penetration for BBBP. The 12-qubit reservoir combines these events before its final $Z+ZZ$ measurement. Increasing $T$ thus exposes more local chemistry while keeping the number of decoder inputs fixed at 78.

At each prefix length, we compare the closed reservoir ($p=0$), the reservoir with prespecified weak reset ($p=0.10$), and a matched 78-node ESN. All three receive the same event streams and use the same scaffold splits. We first compare classification performance using RBF-SVR, then use an affine logistic head to examine the role of nonlinear decoding. The decoder-selection protocol is the same for the quantum and classical reservoirs within each comparison.

\begin{figure}[!htbp]
\centering
\includegraphics[width=\columnwidth]{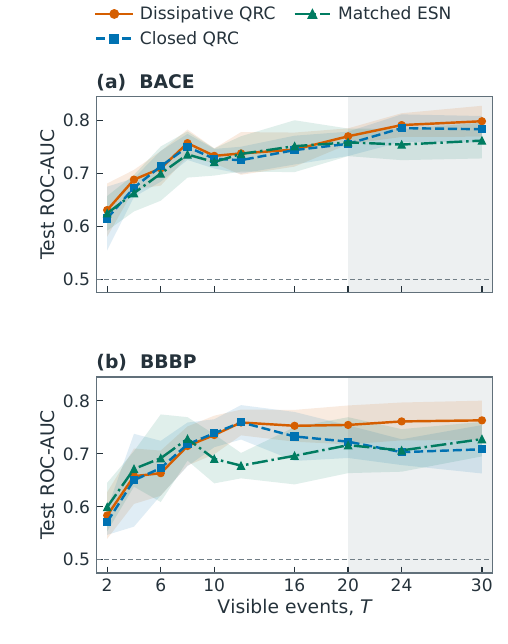}
\caption{Molecular graph classification at matched event length $T$. (a) BACE and (b) BBBP at $Q=12$ under the local chemical-prior token protocol and RBF-SVR decoding. Lines and bands show six paired-seed means and sample standard deviations. Shading marks prefixes with $T>20$.}
\label{fig:graph_main}
\end{figure}

The clearest advantage over the matched ESN occurs on the long prefixes in Fig.~\ref{fig:graph_main}. At $T=30$, weak-reset DTC-QRC reaches a mean BACE AUC of 0.798, compared with 0.762 for the ESN. On both BACE and BBBP, all six paired seeds favor DTC-QRC at $T=24$ and 30 (Table~\ref{tab:graph}). Weak reset also improves the mean AUC over the closed reservoir at the longest prefix on both datasets. With nonlinear decoding, the reservoir can therefore use a long sequence of local events even when the more distant ones receive less weight.

\begin{table}[!htbp]
\caption{Long-prefix molecular classification at matched event length $T$. Values are test ROC-AUC mean $\pm$ sample standard deviation over six paired seeds. Pointwise confidence intervals apply to dissipative DTC-QRC minus matched ESN.}
\label{tab:graph}
\centering
\begingroup
\scriptsize
\setlength{\tabcolsep}{2.8pt}
\renewcommand{\arraystretch}{1.12}
\begin{tabular*}{\columnwidth}{@{\extracolsep{\fill}}lcccc@{}}
\toprule
Task & $T$ & \shortstack{Dissip.\\DTC-QRC} & \shortstack{Matched\\ESN} & \shortstack{$\Delta$ AUC\\\mbox{[95\% CI]}} \\
\midrule
\multirow{2}{*}{BACE} & 24 & $0.791\pm0.022$ & $0.754\pm0.029$ & $0.0368\,[0.0180,\,0.0589]$ \\
 & 30 & $0.798\pm0.029$ & $0.762\pm0.034$ & $0.0366\,[0.0246,\,0.0503]$ \\
\addlinespace[1pt]
\multirow{2}{*}{BBBP} & 24 & $0.761\pm0.036$ & $0.706\pm0.040$ & $0.0547\,[0.0424,\,0.0658]$ \\
 & 30 & $0.763\pm0.037$ & $0.727\pm0.033$ & $0.0355\,[0.0147,\,0.0579]$ \\
\bottomrule
\end{tabular*}
\endgroup
\end{table}

The decoder comparison shows why the amount of retained history alone is insufficient to predict classification performance. At $T=30$ and $p=0.10$, changing the head from affine to RBF-SVR raises BACE AUC from 0.656 to 0.798. BBBP shows the same trend. On both tasks, switching from $p=0$ to weak reset slightly lowers AUC with the affine head but raises it with RBF-SVR. The same change in quantum dynamics thus has different consequences for a linear score and for a decoder that can combine measured coordinates nonlinearly.

\section{Electronic-gap prediction}

In a surface-hopping trajectory, the order of the inputs is set by nuclear motion. We use successive interatomic distances and radial velocities from SHNITSEL A01 ethene trajectories to predict how the electronic gap evolves beyond the observation window. Final-state $Z+ZZ$ measurements supply the inputs to a decoder for each future gap or gap minimum. We assign whole trajectories to the training, validation, and test sets, so every test forecast concerns motion on a trajectory unseen during fitting and reset selection.

\begin{figure}[!t]
\centering
\includegraphics[width=\columnwidth]{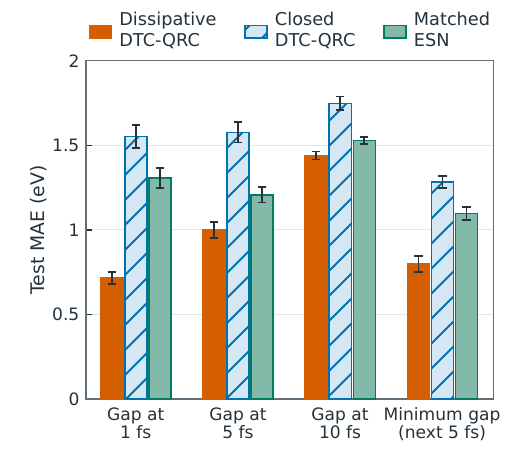}
\caption{Ethene electronic-gap forecasting with validation-selected reset at $Q=15$ and $T=20$. Bars report test MAE for dissipative DTC-QRC, closed DTC-QRC, and the matched ESN using final-state $Z+ZZ$ sampling and RBF-SVR. Error bars show six-seed sample standard deviations.}
\label{fig:hopping_main}
\end{figure}

With final-state sampling and RBF-SVR, DTC-QRC at the reset probability selected on the validation trajectories has lower test MAE than both the closed reservoir and the matched ESN for all four targets (Fig.~\ref{fig:hopping_main}). The largest improvement over the ESN is 0.591 eV for the 1-fs gap. The gain is smaller for the 10-fs gap. For each target the paired 95\% interval excludes zero (Table~\ref{tab:hopping}). The input window and the measured width are fixed throughout. The selected reset probabilities assign different weights to earlier motion at different forecast horizons.

\begin{table}[!htbp]
\caption{Surface-hopping test results after validation-only selection of $p$. MAEs report mean $\pm$ sample standard deviation over six seeds. Intervals use crossed seed-by-trajectory 95\% resampling for DTC-QRC minus matched ESN.}
\label{tab:hopping}
\centering
\begingroup
\scriptsize
\setlength{\tabcolsep}{2.0pt}
\renewcommand{\arraystretch}{1.08}
\begin{tabular*}{\columnwidth}{@{\extracolsep{\fill}}lccc@{}}
\toprule
\shortstack[l]{Target\\selected $p$} & \shortstack{Dissip.\\DTC-QRC (eV)} & \shortstack{Matched\\ESN (eV)} & \shortstack{$\Delta$ MAE (eV)\\\mbox{[95\% CI]}} \\
\midrule
\begin{tabular}[c]{@{}l@{}}Gap, 1 fs\\$p=0.90$\end{tabular} & $0.716\pm0.034$ & $1.306\pm0.057$ & \begin{tabular}[c]{@{}c@{}}$-0.591$\\$[-0.688,\,-0.495]$\end{tabular} \\
\begin{tabular}[c]{@{}l@{}}Gap, 5 fs\\$p=0.65$\end{tabular} & $1.000\pm0.047$ & $1.207\pm0.045$ & \begin{tabular}[c]{@{}c@{}}$-0.207$\\$[-0.286,\,-0.123]$\end{tabular} \\
\begin{tabular}[c]{@{}l@{}}Gap, 10 fs\\$p=0.80$\end{tabular} & $1.440\pm0.024$ & $1.529\pm0.020$ & \begin{tabular}[c]{@{}c@{}}$-0.089$\\$[-0.163,\,-0.015]$\end{tabular} \\
\begin{tabular}[c]{@{}l@{}}Next-5-fs Min.\\$p=0.85$\end{tabular} & $0.800\pm0.047$ & $1.097\pm0.040$ & \begin{tabular}[c]{@{}c@{}}$-0.298$\\$[-0.379,\,-0.213]$\end{tabular} \\
\bottomrule
\end{tabular*}
\endgroup
\end{table}

To test whether the gain over the ESN extends to a different classical predictor, we also fit a tree ensemble to the same inputs. For the 10-fs gap, an Extra Trees regressor yields a test MAE of 1.404 eV, compared with 1.440 eV for DTC-QRC. Thus a classical model operating directly on the input window can outperform the reservoir at this horizon even though the matched ESN has higher error.

\begin{figure*}[!t]
\centering
\includegraphics[width=0.98\textwidth]{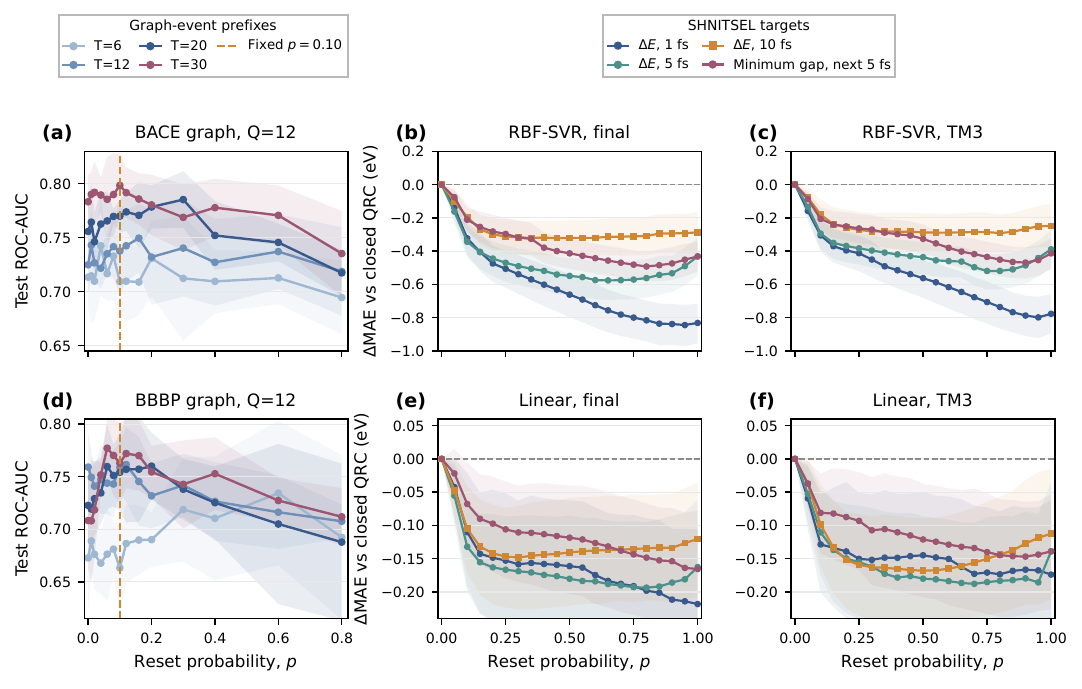}
\caption{Dependence on reset probability for the two molecular tasks. (a,d) BACE and BBBP test AUC at $Q=12$ for four event lengths over the scanned range $0\leq p\leq0.8$. Lines and bands show six-seed means and sample standard deviations. The vertical dashed line marks the prespecified graph setting $p=0.10$. (b,c) Ethene gap forecasts with RBF-SVR and final-state or TM3 sampling. (e,f) The same forecasts with linear decoding. The trajectory panels show the change in test MAE relative to the paired closed reservoir ($p=0$) over the uniform grid $p=0,0.05,\ldots,1$. Negative values favor reset. Shading gives crossed seed-by-trajectory 95\% intervals.}
\label{fig:reset_scans}
\end{figure*}

\begin{figure*}[!t]
\centering
\includegraphics[width=0.98\textwidth]{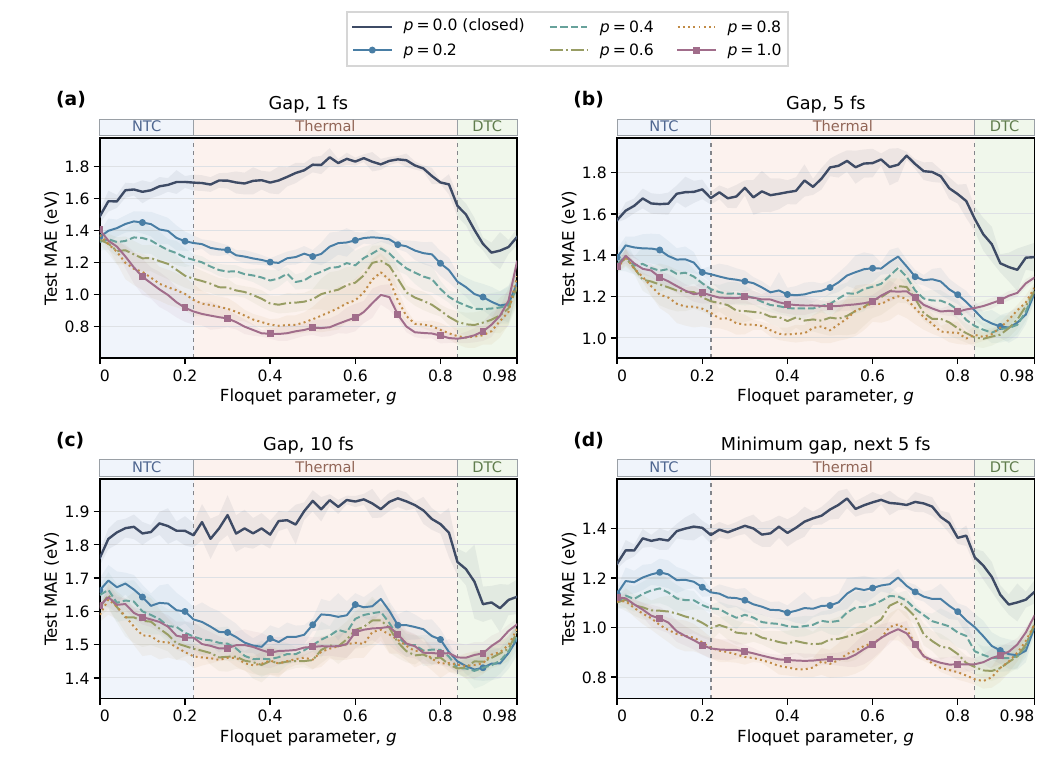}
\caption{Floquet-drive dependence of ethene gap prediction with RBF-SVR and final-state readout. (a)--(d) The 1-, 5-, and 10-fs gaps and the minimum gap in the next 5 fs. Each panel shows $p=0,0.2,\ldots,1$ at $Q=15$, $T=20$, $L=3$, and $W=0.50$. Lines and bands give means and sample standard deviations across six reservoir seeds on one fixed whole-trajectory split. The background labels mark the nontime-crystalline (NTC), thermal, and DTC reference regions from the literature, with divisions at $g=0.22$ and 0.84 \cite{ZhangLiGuoYuJinYin2025,Mi2022Nature}.}
\label{fig:gscan_final}
\end{figure*}

Earlier inputs have different meanings in the two tasks. A graph event can supply information about a group elsewhere in the same molecule, even when the relevant groups enter the sequence many events apart. A trajectory frame describes a particular time in the molecule's motion, and its relevance can change with the forecast horizon.

\section{Effects of controlled dissipation}

\label{sec:dissipation}

Varying $p$ at fixed Floquet parameters changes the weights of input histories of different lengths [Eq.~(\ref{eq:suffix_mixture})]. We use this dependence to ask how much of the graph prefix or trajectory window the final measurement needs to retain. Figure~\ref{fig:reset_scans} compares the two tasks as the reset probability increases.

For graphs, weak reset improves several of the longer prefixes, with the clearest gains on BBBP [Fig.~\ref{fig:reset_scans}(a,d)]. On short prefixes its effect is small or slightly unfavorable. At stronger reset, histories spanning widely separated events carry little weight, and classification can deteriorate. The long-prefix results favor partial forgetting while enough structural context still reaches the measurement.

For gap forecasting, the broad minima of the RBF-SVR curves show that recent motion can support accurate forecasts over a range of reset probabilities [Fig.~\ref{fig:reset_scans}(b,c,e,f)]. The 1-fs gap favors strong reset. For longer horizons the error rises again as reset approaches completeness. The changes are smaller with ridge decoding, and TM3 alters their magnitude more than the location of the low-error regions. The dependence on the prediction target is visible with both decoders and both sampling schemes.

Complete reset provides a direct comparison with a reservoir that retains no state between encoded frames. At $p=1$, $\rho_T=\mathcal{F}_{x_T}(\rhozero)$, so the final state depends only on the last event and its subsequent evolution. That event still contains distances and radial velocities, with each velocity computed from two adjacent raw frames. Relative to this local description of motion, retaining reservoir history barely changes the mean 1-fs error but lowers the 5-fs error by about 0.14 eV (Table~\ref{tab:hopping_p1}). Earlier encoded frames therefore add predictive value beyond the final distance/velocity pair, with the largest gain at the intermediate horizon.

\begin{table}[!htbp]
\caption{Complete-reset comparison for surface hopping at $Q=15$, $T=20$, with final-state $Z+ZZ$ and RBF-SVR. The $p=1$ column gives six-seed mean test MAE $\pm$ sample standard deviation. The last column gives the mean difference $\mathrm{MAE}(1)-\mathrm{MAE}(p_\star)$, where Table~\ref{tab:hopping} reports the validation-selected $p_\star$. Positive differences favor retaining reservoir history.}
\label{tab:hopping_p1}
\centering
\begingroup
\scriptsize
\setlength{\tabcolsep}{3pt}
\renewcommand{\arraystretch}{1.12}
\begin{tabular*}{\columnwidth}{@{\extracolsep{\fill}}lcc@{}}
\toprule
Target & $p=1$ MAE (eV) & $\mathrm{MAE}(1)-\mathrm{MAE}(p_\star)$ (eV) \\
\midrule
Gap, 1 fs & $0.722\pm0.027$ & $+0.0065$ \\
Gap, 5 fs & $1.144\pm0.018$ & $+0.1442$ \\
Gap, 10 fs & $1.461\pm0.044$ & $+0.0217$ \\
Next-5-fs Min. & $0.852\pm0.026$ & $+0.0527$ \\
\bottomrule
\end{tabular*}
\endgroup
\end{table}

The 5- and 10-fs forecasts illustrate why a longer horizon need not favor a longer reservoir memory. Both benefit on average from retaining earlier frames relative to complete reset. Yet the 10-fs target selects stronger reset than the 5-fs target and has higher test error. A possible explanation is that some of the motion in earlier frames repeats information already present in the most recent observations, while other motion contributes little to the target. In a readout of fixed width, giving all of this history more weight can make the relevant motion harder to distinguish.

The factor $(1-p)^\ell$ determines the probability that a history survives across $\ell$ reset locations. For any history that does survive, the Floquet drive determines how its local inputs combine before measurement. The reset probability $p$ sets the weight of each history, and the drive parameter $g$ shapes how the retained inputs appear in the final measurements used by the decoder.

\section{Floquet-drive dependence}

\label{sec:gscan}

Can one value of $g$ keep forecast errors low for targets that benefit from different amounts of past motion? We scan $g$ at fixed reset probabilities, keeping the ethene input window, reservoir size, trajectory split, and seeds the same. Validation selects the decoder hyperparameters at each setting. Figure~\ref{fig:gscan_final} shows the resulting test errors for the four targets with RBF-SVR and final-state $Z+ZZ$ readout.

To compare the 20 combinations of target and nonzero reset probability, we express each six-seed mean test MAE in terms of its percentage excess over the minimum on its own drive curve. The equal-weight mean excess is smallest at $g=0.88$. Ranking the drives within each configuration instead gives the best mean rank at $g=0.86$. Both summaries of the nonmonotonic curves therefore locate a common low-error window at $g\simeq0.86$--$0.88$.

This window lies close to the DTC transition reference $g_c\simeq0.84$ for related Floquet spin chains \cite{Mi2022Nature}. At the fixed application setting $g=0.84$, 17 of the 20 dissipative configurations with RBF-SVR and final-state readout have a mean MAE within $5\%$ of their own scanned minimum. The transition reference therefore gives a practical fixed drive across most of these forecasts.

At $p=1$ the final frame enters a freshly reset reservoir. Even then, the 1-fs MAE with RBF-SVR and final-state readout falls from 1.403 eV at $g=0$ to 0.722 eV at $g=0.84$. With no state retained between frames, this difference comes from how the drive maps the last distances and radial velocities into measurable features. Appendix~\ref{app:gscan} examines how the decoder and sampling times affect the preferred drive.

\section{Dephasing effects}

\begin{figure}[!htbp]
\centering
\includegraphics[width=\columnwidth]{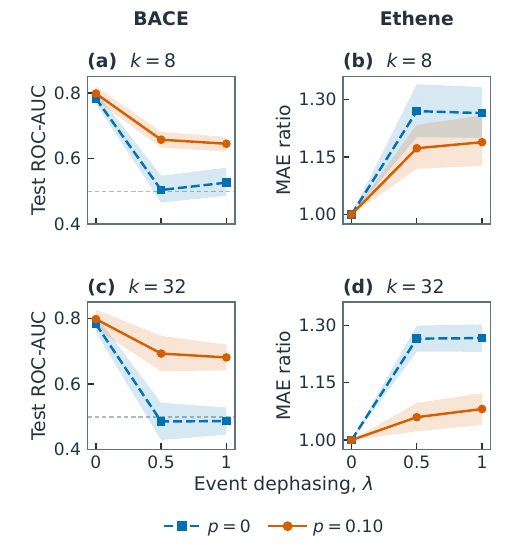}
\caption{Event-wise dephasing with $k=8$ Pauli trajectories in (a,b) and $k=32$ in (c,d). Left panels show BACE test ROC-AUC ($Q=12$, $T=30$). Right panels show target-balanced gap MAE relative to the coherent model at the same $p$ ($Q=15$, $T=20$). Both use final-state $Z+ZZ$ and RBF-SVR at $p=0$ or 0.10, with exact evolution at $\lambda=0$. Lines give six-seed means. Bands at $k=8$ are 95\% bootstrap intervals over paired seeds for BACE and crossed seeds and trajectories for ethene. Bands at $k=32$ give sample standard deviations of seed means.}
\label{fig:dephasing}
\end{figure}

To test the role of coherent propagation, we apply dephasing after each event at fixed $p$. BACE AUC decreases at both Pauli-trajectory counts [Fig.~\ref{fig:dephasing}(a,c)]. At $\lambda=0.5$ and $k=8$, for example, the closed-reservoir AUC falls from 0.783 to 0.504. Weak reset reduces the loss at both counts.

Gap forecasting shows the same loss of accuracy [Fig.~\ref{fig:dephasing}(b,d)]. For both reset settings and trajectory counts, the target-balanced MAE ratio exceeds one relative to coherent evolution at the same $p$.

Dephasing in the computational basis leaves the instantaneous $Z$ and $ZZ$ expectations unchanged. Later input rotations and Floquet cycles, however, can convert coherence into populations and pair correlations. Suppressing coherence after an event therefore changes what later measurements can reveal about that input. The observed losses are consistent with a role for coherent propagation in both tasks. They reflect the full action of the channel, including its effect on purity and subsequent mixing.

\section{Experimental demonstration}

\begin{figure}[!htbp]
\centering
\includegraphics[width=\columnwidth]{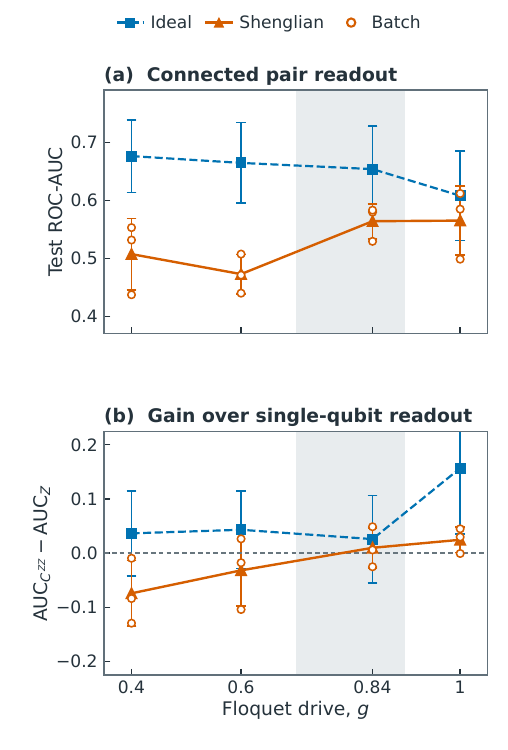}
\caption{Four-point drive scan on the Quafu superconducting quantum cloud platform. (a) Test ROC-AUC from the 15 connected pair coordinates $C_{ij}^{ZZ}=\langle Z_iZ_j\rangle-\langle Z_i\rangle\langle Z_j\rangle$ available at $Q=6$. (b) Connected-$ZZ$ AUC minus the separately fitted six-dimensional $Z$-only AUC. Ideal markers show means and sample standard deviations over ten nested scaffold/decoder analyses. Hardware markers show means and sample standard deviations over three Shenglian acquisition batches. Open circles show individual batch means. Every point uses the same balanced 400-molecule BACE subset, $Q=6$, $T=6$, $L=3$, $W=0.6$, $p=0$, coupling realization, 180 nominal controlled-$X$ (CX) gates, 1024 shots, and affine decoder protocol. The gray region marks the $g=0.7$--0.9 transition-edge window inherited from prior work.}
\label{fig:hardware_g}
\end{figure}

In simulation, varying the reset probability reveals when retaining earlier inputs improves prediction. For molecular classification on a noisy processor, the question is which low-order observables still distinguish active from inactive compounds. We run the BACE classifier on the Baihua and Shenglian processors of the Quafu superconducting quantum cloud platform. In the closed protocol ($p=0$), six-qubit circuits encode successive graph events with $U_3$ gates and evolve between events without a programmed reset. An affine decoder assigns a molecular activity score from the measured $Z+ZZ$ coordinates. Appendix~\ref{app:hardware} gives the input-angle mapping, device characteristics, and acquisition protocol.

With this closed protocol on Shenglian, we compare the complete classifier at four drives spanning the thermal region and the DTC side. For the same balanced BACE subset, the full $Z+ZZ$ readout reaches its highest mean test AUC at $g=0.84$, with a value of 0.585 across three acquisition batches. To understand why the transition edge is favorable, we use the same measurements to examine what the pair correlations contribute to classification.

Subtracting the product of the two local polarizations gives the connected correlation for each qubit pair. With these correlations alone, ideal AUC decreases across the four drives as $g$ increases. Hardware AUC follows a different pattern. It is higher at the two drives toward the DTC side and is comparable at the edge and deeper in the DTC region, even though the ideal AUC is higher at the edge [Fig.~\ref{fig:hardware_g}(a)]. The connected-pair decoder also outperforms the $Z$-only decoder at the two higher drives, reversing their ordering at the lower drives [Fig.~\ref{fig:hardware_g}(b)].

We fit the affine head separately to the ideal and measured coordinates. A smaller drop in AUC therefore means that the noisy pair observables retain more of their ability to distinguish active from inactive molecules. The depth control gives a second indication of their sensitivity to noise. Folding the circuit produces the largest Shenglian AUC loss in the pairwise readout (Appendix~\ref{app:hardware}). The three $g=0.84$ batches were acquired one day before the interleaved runs at the other drives, so variation across days can also contribute to the observed drive dependence.

The drive dependence suggests a balance between input mixing and robustness to device noise. In related Floquet systems, thermalizing dynamics spread local perturbations more rapidly, while DTC dynamics restrict their propagation \cite{Mi2022Nature,Ippoliti2021PRXQ}. Mixing can bring different molecular inputs into the same observable. Toward the DTC side, the smaller loss in pair-based classification indicates that the measured correlations retain more of their predictive value. Near the transition, the ideal AUC from connected pair correlations is higher than deeper in the DTC region, and the drop in AUC on hardware is smaller than on the thermal side. This balance offers a physical explanation for the maximum in full-readout classification near the transition.

\section{Conclusion and outlook}

The same rules for Floquet evolution and measurement support molecular property prediction from both structure and motion. Extending the single-encoding DTC-QRC image classifier \cite{ZhangLiGuoYuJinYin2025} to successive molecular events introduces a choice of how long a previous observation should contribute. Global reset makes that choice explicit. It weights histories of different lengths while leaving the number of observables required by the final readout fixed. In simulations with nonlinear decoding and matched input lengths and readout widths, DTC-QRC outperforms the ESN both in classifying molecules from long BACE and BBBP graph prefixes and in forecasting all four ethene gap targets. Dephasing lowers accuracy in both molecular applications, consistent with a role for coherent propagation.

The two molecular applications favor different reset probabilities. Weak reset leaves enough chemical context to classify a molecule from its local graph events. The gap forecasts favor recent motion, but the benefit of retaining earlier frames varies with the forecast horizon. Relative to complete reset, retaining earlier frames gives its largest mean error reduction for the 5-fs gap. For a given decoder, the useful memory depends on what the target requires beyond the current observation.

For a given input history, the drive shapes the molecular information available to the decoder. Even at complete reset, the accuracy of the gap forecast from the final encoded frame depends on the drive. For dissipative ethene forecasts with RBF-SVR and final-state readout, the targets share a low-error window just above the DTC transition reference for related spin chains \cite{Mi2022Nature}. The BACE experiment on the Quafu superconducting quantum cloud platform holds reset at $p=0$ and tests which observables retain predictive value as the drive varies. Of the four drives sampled on Shenglian, the one near the transition gives the highest mean AUC for the full $Z+ZZ$ readout. Toward the DTC side, the AUC obtained from connected pair correlations stays closer to its ideal value. Near the transition, input mixing can bring several local events into the same observable while pair correlations retain their ability to distinguish molecules. These observations favor the transition edge as a starting point for drive selection on a noisy processor.

For a new molecular target, the gain from retaining earlier input events can guide the choice of reset probability in simulation. Comparing candidate drives and readouts on the intended processor would then identify which measured correlations best predict the property under device noise. For molecular screening, a stream of local contacts around a binding site could extend this approach to affinity prediction. Along a longer conformational trajectory, the same architecture could weight motion on the timescale of an electronic response.

\begin{acknowledgments}
The authors thank Haifeng Yu and Quafu team at the Beijing Academy of Quantum Information Sciences for supporting the superconducting quantum cloud experiments. This work is supported by Beijing Institute of Technology Research Fund Program under Grant No.~2024CX01015, the Fundamental Research Funds for the Central Universities, and the National Natural Science Foundation of China under Grant No.~62533015.

The authors used GPT-5.6 SOL and GPT-6 Astra for literature search and synthesis, manuscript revision, language editing, and proofreading. The authors reviewed the manuscript and take responsibility for its final content.
\end{acknowledgments}

\setlength{\textfloatsep}{8pt plus 2pt minus 2pt}
\setlength{\floatsep}{8pt plus 2pt minus 2pt}
\setlength{\dbltextfloatsep}{8pt plus 2pt minus 2pt}

\section*{Data availability}

The BACE and BBBP tasks follow MoleculeNet \cite{Wu2018MoleculeNet}. SHNITSEL A01 is available from the repository described in Ref.~\cite{Curth2025SHNITSEL}. The \href{https://github.com/eternal-fly/article-git\#qrc-quantum-reservoir-computing}{GitHub repository} contains the code used to simulate the reservoirs, run the classical benchmarks, analyze the Quafu measurements, and generate the figures \cite{DTCQRCCode2026}. The raw Quafu measurement counts and numerical results for individual seeds are available from the corresponding author upon reasonable request.

\appendix
\renewcommand{\appendixname}{APPENDIX}
\makeatletter
\setlength{\@fptop}{0pt}
\setlength{\@fpsep}{12pt}
\setlength{\@dblfptop}{0pt}
\setlength{\@dblfpsep}{12pt}
\makeatother

\section{SUPERCONDUCTING DEVICES AND CONTROL EXPERIMENTS}
\label{app:hardware}

The Baihua and Shenglian processors provide complementary tests of molecular classification under device noise. Table~\ref{tab:devices} summarizes their reported characteristics and the six-qubit paths used in the experiments \cite{Liu2025Baihua,BAQIS2025BaihuaCharacterization,Zhang2026ShenglianCalibration}. Both devices couple transmon qubits through tunable couplers and support native controlled-$Z$ (CZ) gates. We submit circuits on the specified physical paths with compilation enabled and readout correction disabled. Each circuit uses 1024 shots. The $L=1$ base and $L=3$ circuits contain 60 and 180 nominal controlled-$X$ (CX) gates, respectively.

\begin{figure}[!htbp]
\centering
\includegraphics[width=\columnwidth]{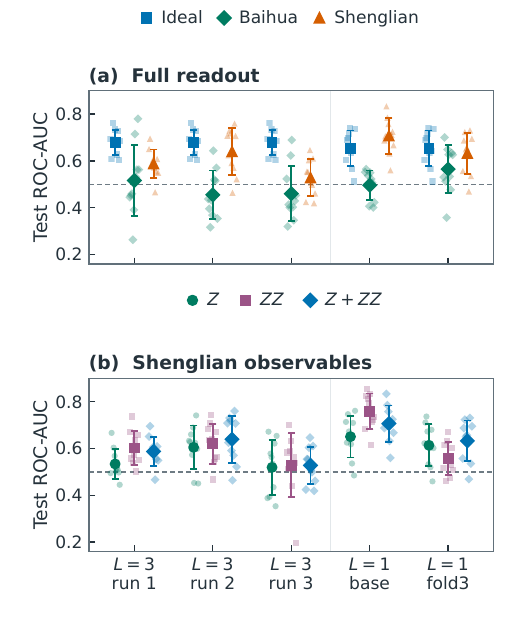}
\caption{Molecular classification and depth controls on the Quafu superconducting quantum cloud platform. (a) Full $Z+ZZ$ test AUC for matched ideal, Baihua, and Shenglian features across three $L=3$ acquisition batches and the paired $L=1$ base/fold3 circuits. (b) Shenglian AUC after retaining $Z$, $ZZ$, or their union. Large markers and bars show means and sample standard deviations over ten nested scaffold/decoder analyses. Pale points show the individual analyses. The ten analyses reuse one feature matrix per hardware condition, while acquisition batches represent device repetitions. Fold3 preserves the ideal $L=1$ unitary while increasing the nominal CX count from 60 to 180. The observable families expose different widths. Dashed lines mark chance.}
\label{fig:hardware_execution_supp}
\end{figure}

\begin{table*}[!t]
\caption{Quafu processors and experimental settings. Coherence times and gate fidelities refer to the published characterization samples and dates. Physical paths and shot counts refer to the present experiment. Shenglian medians cover 65 qubits and 72 CZ gates. The Baihua CZ mean covers 130 gates. The $T_2$ measurements use spin echo for Baihua and Carr--Purcell--Meiboom--Gill (CPMG) refocusing for Shenglian.}
\label{tab:devices}
\centering
\begingroup
\footnotesize
\setlength{\tabcolsep}{6pt}
\renewcommand{\arraystretch}{1.0}
\begin{tabular*}{\textwidth}{@{\extracolsep{\fill}}lll@{}}
\toprule
Parameter & Baihua \cite{Liu2025Baihua,BAQIS2025BaihuaCharacterization} & Shenglian \cite{Zhang2026ShenglianCalibration} \\
\midrule
Physical qubits & 156 & 84 \\
Qubit type & Fixed-frequency transmon & Frequency-tunable transmon \\
Connectivity & Heavy-hexagon-like & Hexagonal \\
$T_1$ ($\mu$s) & 77 (mean) & 50.8 (median) \\
$T_2$ ($\mu$s) & 58 (spin echo, mean) & 16.9 (CPMG, median) \\
Single-qubit gate fidelity & $>99.9\%$ (mean) & $99.95\%$ (median) \\
CZ gate fidelity & $98.65\%$ (mean) & $99.25\%$ (median) \\
\midrule
Physical qubit path & $72,73,74,75,76,77$ & $27,34,41,48,54,61$ \\
Path-selection map date & 28 July 2026 & 28 July 2026 \\
Shots per circuit & 1024 & 1024 \\
Readout correction & Disabled & Disabled \\
\bottomrule
\end{tabular*}
\endgroup
\end{table*}

Figure~\ref{fig:hardware_execution_supp} summarizes the execution controls that complement the four-point $g$ scan in the main text. The canonical $Q=6$, $T=6$ circuits use the same label-free local chemical-prior token rule on a balanced 400-molecule BACE subset. For each event, we apply $\tanh$ elementwise to its $3Q$ token coordinates and assign successive triples to the $U_3$ angles $(\theta,\varphi,\chi)$ on each qubit. We supply the angles in radians. Up to a global phase, $U_3(\theta,\varphi,\chi)=R_z(\varphi)R_y(\theta)R_z(\chi)$. We measure every molecular circuit with 1024 shots, and one computational-basis count table supplies all six $Z$ and 15 $ZZ$ coordinates. We refit the same affine decoder with a bias separately to each set of ideal or hardware features.

For the Shenglian four-point drive comparison in Fig.~\ref{fig:hardware_g}, three acquisition rounds interleaved the $g=0.40,0.60,$ and 1.00 circuits. The three $g=0.84$ batches were acquired one day earlier. All settings share the molecular subset, qubit chain, depth, measurement basis, nominal two-qubit-gate count, and decoder protocol.

The $L=3$ comparison tests end-to-end classification on both processors. Matched ideal $Z+ZZ$ features give test AUC $0.679\pm0.054$. The three Shenglian batches give $0.587\pm0.061$, $0.640\pm0.100$, and $0.528\pm0.079$, whereas the corresponding Baihua values are $0.517\pm0.151$, $0.455\pm0.104$, and $0.460\pm0.118$. Each uncertainty is the sample standard deviation across ten nested scaffold/decoder analyses of a fixed feature matrix. Acquisition batches are separate device repetitions, while the ten analyses quantify variability in splitting and decoding the same measurements. All three Shenglian batch means exceed 0.5, showing that the measured observables retain information useful for classification. Differences across batches and processors reflect acquisition-dependent changes in that representation.

The depth control changes the implemented gate sequence while preserving the ideal map. The base circuit implements the $L=1$ unitary $B$, and fold3 inserts $B^{\dagger}B$, giving $BB^{\dagger}B=B$ without noise. Both circuits therefore have the same ideal $Z+ZZ$ AUC, $0.654\pm0.076$. Increasing the nominal CX count from 60 to 180 changes Shenglian AUC from $0.707\pm0.078$ to $0.633\pm0.087$, a nested-analysis mean decrease of $0.074$. Baihua changes from $0.496\pm0.063$ to $0.565\pm0.102$, an increase of $0.069$. The opposite AUC shifts indicate that folding changes how well the measured features distinguish the two classes on each processor, beyond the effect of uniform attenuation.

Resolving the Shenglian readout by observable family shows where folding has its largest effect. The $ZZ$-only AUC falls from $0.759$ to $0.556$, and the loss appears in all ten nested analyses. The corresponding decreases for $Z$ alone and for the full readout are $0.037$ and $0.074$. The pair observables are therefore the most sensitive to added depth in this comparison.

\section{ESN WIDTH DEPENDENCE}
\label{app:esn_width}

The main comparison matches the ESN readout width to that of QRC. Here we examine how increasing ESN width changes molecular classification, using 78, 500, 1000, and 2000 recurrent nodes on BACE and BBBP at $T=12$ and 30.

\begin{figure}[!ht]
\centering
\includegraphics[width=\columnwidth]{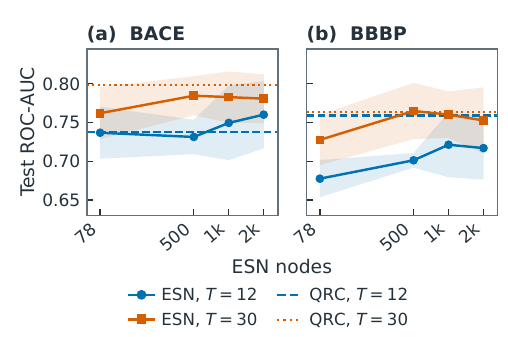}
\caption{ESN width dependence at fixed molecular input. Test ROC-AUC on (a) BACE and (b) BBBP for $T=12$ and 30. Solid curves and shading give ESN means and sample standard deviations over six scaffold-split seeds. Horizontal dashed and dotted lines give the corresponding QRC means at $p=0.10$. Each ESN width uses its own validation-selected decoder.}
\label{fig:esn_width}
\end{figure}

\begin{samepage}
All widths use the local chemical-prior tokens and the same six scaffold splits as the matched comparison. The quantum reference fixes $Q=12$, $g=0.84$, $W=0.50$, $L=3$, and $p=0.10$, with 78 final-state $Z+ZZ$ coordinates. The ESNs retain spectral radius 0.90, input scale 0.75, leak rate 0.65, and three updates per event.

\end{samepage}

Each width uses the graph RBF-SVR grid in Appendix~\ref{app:protocol} and training-feature standardization. We select decoder hyperparameters separately for each dataset, event length, width, and seed using the same validation protocol.

Increasing the ESN width does not consistently improve test AUC (Fig.~\ref{fig:esn_width}). At $T=30$, both datasets reach their highest mean ESN AUC among the widths tested at 500 nodes. For BBBP, this gives $0.765$, close to $0.763$ for QRC, and the mean falls to $0.752$ at 2000 nodes. We interpret the near agreement on BBBP as reflecting a weaker dependence on graph traversal order than for BACE. If the relevant local chemistry is less sensitive to the order in which the predictor encounters it, a wider ESN can capture enough of that chemistry to match QRC at this prefix length.

\begin{figure*}[!t]
\centering
\includegraphics[width=0.98\textwidth]{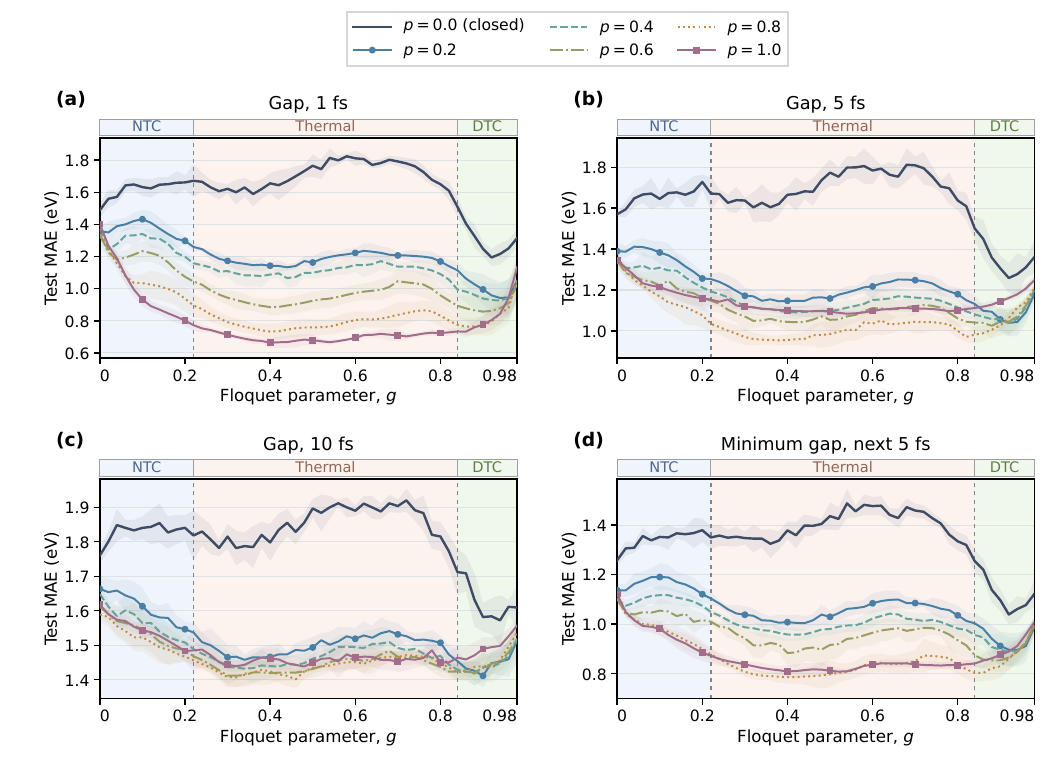}
\caption{Floquet-drive dependence with RBF-SVR and TM3 readout. Panels (a)--(d) follow the target order in Fig.~\ref{fig:gscan_final}. All six reset values, reservoir parameters, seeds, and data partitions match that comparison. TM3 concatenates three post-input $Z+ZZ$ samples. Lines and bands show six-seed means and sample standard deviations. Background regions retain the same literature reference divisions as Fig.~\ref{fig:gscan_final}.}
\label{fig:gscan_tm3}
\end{figure*}

\begin{table*}[!t]
\caption{Drive windows across decoder and readout choices. $N$ counts target--reset configurations. The columns $g_E$ and $g_R$ minimize equal-weight mean percentage excess MAE and mean rank, respectively, over the scanned drives. $E_{\min}$ gives the minimum mean excess. The last column counts configurations with mean MAE within $5\%$ of their own scanned minimum at the fixed drive $g=0.84$. The rows marked $p>0$ include $p=0.2,0.4,0.6,0.8,1$.}
\label{tab:gscan_windows}
\centering
\begingroup
\footnotesize
\setlength{\tabcolsep}{5pt}
\renewcommand{\arraystretch}{1.04}
\begin{tabular*}{\textwidth}{@{\extracolsep{\fill}}llccccc@{}}
\toprule
Decoder/readout & Reset set & $N$ & $g_E$ & $E_{\min}$ (\%) & $g_R$ & Within $5\%$ at 0.84 \\
\midrule
RBF-SVR, final & All $p$ & 24 & 0.90 & 2.50 & 0.86 & 17/24 \\
RBF-SVR, final & $p>0$ & 20 & 0.88 & 1.61 & 0.86 & 17/20 \\
\addlinespace[2pt]
RBF-SVR, TM3 & All $p$ & 24 & 0.92 & 3.64 & 0.86 & 13/24 \\
RBF-SVR, TM3 & $p>0$ & 20 & 0.90 & 3.67 & 0.42 & 13/20 \\
\addlinespace[2pt]
Ridge, final & All $p$ & 24 & 0.90 & 1.02 & 0.88 & 21/24 \\
Ridge, final & $p>0$ & 20 & 0.88 & 0.93 & 0.88 & 20/20 \\
\addlinespace[2pt]
Ridge, TM3 & All $p$ & 24 & 0.92 & 1.34 & 0.44 & 20/24 \\
Ridge, TM3 & $p>0$ & 20 & 0.44 & 0.99 & 0.44 & 20/20 \\
\addlinespace[2pt]
All decoders/readouts & All $p$ & 96 & 0.90 & 2.17 & 0.86 & 71/96 \\
All decoders/readouts & $p>0$ & 80 & 0.88 & 1.98 & 0.86 & 70/80 \\
\addlinespace[2pt]
\bottomrule
\end{tabular*}
\endgroup
\end{table*}

\section{DRIVE DEPENDENCE ACROSS DECODERS AND READOUTS}

\label{app:gscan}

The main-text drive window uses the primary RBF-SVR decoder and final-state readout. Here we examine its dependence on decoding and sampling by including linear ridge regression and TM3. Final sampling gives 120 observables, while TM3 gives 360 from three post-input sampling times. Four targets, six reset values, two decoders, and two readouts yield 96 configurations, of which 80 have nonzero reset. All use the drive grid $g=0,0.02,\ldots,0.98$. Figure~\ref{fig:gscan_tm3} shows RBF-SVR with TM3, and Table~\ref{tab:gscan_windows} compares the preferred drives across both decoders and readouts.

For each configuration we rank the 50 drives by six-seed mean test MAE, assigning rank 1 to the lowest error. We also calculate the percentage excess above that minimum as in Sec.~\ref{sec:gscan}. Equal-weight averages over configurations summarize the fixed-setting test curves. In the main application comparisons the physical settings and the validation procedure for choosing the decoder are as specified in the methods.

A favorable high-$g$ window survives the combined comparison (Table~\ref{tab:gscan_windows}). Across all configurations, minimum mean excess and best mean rank place it at $g\simeq0.86$--$0.90$. Across the dissipative settings the corresponding window is $g\simeq0.86$--$0.88$. This agreement supports the shared high-$g$ region as a useful initial choice across the tested targets and reservoir settings.

TM3 also reveals a low-$g$ region. With dissipation and ridge decoding, both criteria favor $g=0.44$. With dissipative RBF-SVR/TM3, the mean rank favors $g=0.42$ but the mean excess favors $g=0.90$. A rank counts the order of the errors, so a narrow win and a large win count alike. The percentage excess also records how far an unfavorable choice lies above the minimum. Measuring after additional Floquet cycles gives the decoder different projections of the encoded state and can change which drive it favors.

Even at complete reset, the preferred drive depends on when the reservoir is measured. For $p=1$ with RBF-SVR/TM3, the four target-specific minima lie between $g=0.40$ and 0.54, whereas the corresponding final-readout mean percentage excess is lowest at $g=0.84$. Even without memory between events, changing the sampling times changes which drive produces the most useful features. Joint selection of the drive and readout can thus refine the shared high-$g$ starting point for a particular molecular target.

\section{DATA PARTITIONS AND MODEL SELECTION}
\label{app:protocol}

The six paired graph splits use seeds $0,10,20,30,40,50$. The BBBP records include 60 canonical structures that occur more than once, ten with conflicting labels. We retain the original labels and group records by structure within each split. The pointwise graph intervals use 20,000 bootstrap draws with random seed 20260905.

For graph RBF-SVR we search $C\in\{0.1,1,10\}$ and median-distance gamma scales $\{0.25,1,4\}$ at $\epsilon=0.01$. We select by highest validation AUC, breaking ties by lower clipped-score log loss and then higher clipped-score accuracy. The logistic head uses binary cross entropy for 500 epochs and the same rule to select an epoch.

For ethene, the ridge penalties are $\alpha\in\{10^{-6},10^{-4},10^{-2},0.1,1,10,100\}$. RBF-SVR uses $C\in\{0.1,1,10,100\}$, gamma scales $\{0.0625,0.25,1,4\}$ divided by the feature dimension, and $\epsilon\in\{0.005,0.02\}$. For reset selection we supplement the 21 values $p=0,0.05,\ldots,1$ with $0.0025,0.005,0.01,0.02,0.04,0.08$.

\bibliography{references}

\end{document}